\documentclass[twoside,twocolumn]{article}

\usepackage{blindtext} 

\usepackage{graphicx}
\usepackage{subfigure}

\usepackage[sc]{mathpazo} 
\usepackage[T1]{fontenc} 
\usepackage{microtype} 

\usepackage[english]{babel} 

\usepackage[hmarginratio=1:1,top=32mm,columnsep=20pt]{geometry} 
\usepackage[hang, small,labelfont=bf,up,textfont=it,up]{caption} 
\usepackage{booktabs} 

\usepackage{lettrine} 

\usepackage{enumitem} 
\setlist[itemize]{noitemsep} 

\usepackage{abstract} 

\usepackage{titlesec} 
\renewcommand\thesection{\Roman{section}} 
\renewcommand\thesubsection{\roman{subsection}} 
\titleformat{\section}[block]{\large\scshape\centering}{\thesection.}{1em}{} 
\titleformat{\subsection}[block]{\large}{\thesubsection.}{1em}{} 

\usepackage{fancyhdr} 
\usepackage{titling} 

\usepackage{hyperref} 

\newcommand{\ep}{\epsilon}

\newcommand{\om}{\omega}

\newcommand{\beq}{\begin{equation}}
\newcommand{\eeq}{\end{equation}}

\newcommand{\ball}{\begin{align}}
\newcommand{\eall}{\end{align}}

\newcommand{\beqar}{\begin{eqnarray}}
\newcommand{\eeqar}{\end{eqnarray}}

\newcommand{\ben}{\begin{enumerate}}
\newcommand{\een}{\end{enumerate}}

\pretitle{\begin{center}\huge\bfseries} 
\posttitle{\end{center}} 
\title{A plausible explanation of the repeated "noise" pattern in the data of arXive:1807.08572: quantum confinement and surface phonon modes} 
\author{%
\textsc{Navinder Singh}\thanks{Cell Phone: +919662680605; Landline: 00917926314457.} \\[1ex] 
\normalsize Physical Research Laboratory, Ahmedabad, India. \\ 
\normalsize \href{mailto:navinder.phy@gmail.com}{navinder.phy@gmail.com} 
}
\date{\today} 
\renewcommand{\maketitlehookd}{%
\begin{abstract}
\noindent  A key point from the experiment related to the  "noise" pattern in figure 3(a) of arXive:1807.08572 is that it is  a function of temperature. We put forward a possible explanation. We argue that it is the elasto-magnetic coupling in which the diamagnetic surface currents on the surface of nanoparticles couple to surface phonon modes. As the temperature is increased  (say from 210 K to 220 K), successive phonon modes are excited, one by one. This is due to quantum confinement where phonon modes take discrete values of energy (in bulk, the modes are quasi-continuous). One by one excitation of phonon modes modulates the surface currents on nano-particles that led to a fluctuating magnetization thus "noise" in the lower part of the susceptibility data where the system is in the superconducting state. 
\end{abstract}
}

\begin{document}

\maketitle


On July 23rd a preprint appeared on the arXive reporting the observations of  room temperature superconductivity in a nanostructure solid (D. K. Thapa and A. Pandey, arXive: 1807.08572). The nano-composite material consists of silver nanoparticles ($\sim 1~nm$) in gold matrix which is roughly of spherical shape with radius about $~10 ~nm$. The films and pellets used in the experiments consist of this nano-composites, and superconductivity at 236 K is reported in figure 3(a) of ref\cite{1}.  However, Brian Skinner\cite{2} pointed out a very curious correlations in the noise (refer to expanded version of the noise part of the data in\cite{2}).

The question whether the noise seen in the susceptibility data is actually a noise or a part of a signal is posed by Pratap Raychaudhuri\cite{3}. He argues that: "The central premise based on which the suspicion of academic misconduct arises is that the "noise" is repeating in independent measurements of susceptibility performed on the same sample in different magnetic field. This premise holds only if one can indeed identify the noise from the signal. But what if the component of the data that is identified as noise is not noise at all but is rather part of the signal"\cite{3}.

He offered a plausible explanation which goes like this. The applied magnetic field induces torque on individual magnetic particles and under this torque these particle tend to move in an irregular manner (just as sand particles move when you push it by hand). He further argues that in a weaker magnetic field (<0.1 T) in the case of figure 3 (a) of\cite{1} nanoparticles do not completely detach from each other. When magnetic field is turned off they regain their original position (some sort of a memory effect).  This could be the reason for repeated "noise" pattern. However, larger magnetic field is able to permanently detach them, and that "memory effect" is lost! (thus repetition appears to be lost at say 1 Tesla and above).

In this note we argue that this explanation has one problem. The particles involved are not magnetic particle. There is no permanent magnetic moment of  nanoparticles! Induced magnetization when the system is in the superconducting state is due to the circulating currents on the boundary of these particles, and it aligns itself opposite to the applied magnetic field. Thus the origin of torque is under question.

But there is another possibility which I would like to point out. The key point from experiment is that the "noise" pattern is a function of temperature (at least at weaker magnetic fields (<0.1 T). So one possible explanation is that diamagnetic surface currents on the surface of nanoparticles are modulated when successive phonon modes are excited, one by one,  as the temperature is varied.  The discrete phonon modes are due to confinement (on the surface of nano-particles) where phonon modes take discrete values of energy (in bulk, the modes are quasi-continuous).  A simple minded calculation can be done as follows:

Let $r$ be the radius of the nano-particle (in Thapa-Pandey experiment they are roughly of 10 nm  in diameter).  Periodic boundary condition demands that longest wavelength mode is given by $\lambda = 2\pi r$, for the longest wavelength surface phonon mode. For $n$th mode $n \lambda_n = 2\pi a$. The corresponding wave vector is $k_n = \frac{2\pi}{\lambda_n} = \frac{n}{a}$. And the difference between two nearby wave vectors is $\Delta k = k_n -k_{n-1} = \frac{1}{a}$

Let $v_s^{eff}$ sound speed on the surface of the nanoparticles. Then the energy of the $n$th phonon mode is given by
\beq
\ep_n^{ph} = \hbar \om_n = \hbar k_n v_s^{eff}.
\eeq
The energy gap between two nearby phonon modes is given $\Delta \ep_n^{ph} =  \hbar \Delta k_n v_s^{eff}$. Corresponding temperature "step" is given as
\beq
\Delta T^{ph} = \frac{\hbar \Delta k v_s^{eff} }{k_B} = \frac{\hbar v_s^{eff}}{a k_B} 
\eeq
For  an order of magnitude calculation taking sound speed ($\sim 10^3 ~meters/sec$) we have:
\beq
\Delta T^{ph} = \frac{\hbar v_s^{eff}}{a k_B}  \simeq \frac{10^{-34} \times 10^3}{10\times 10^{-9} \times 10^{-23}} \simeq 1~K
\eeq
This is roughly the spacing between the oscillatory features of the "noise" in the data in figure 3(a) of the original paper! A better calculation can be done\cite{4}, however, the conclusion will not be altered.

So the mechanism of the "noise" might be the successive excitations of surface phonon modes (one by one) as the temperature is increased.  Due to magneto-elastic coupling, these excitations of the phonon modes modulate the magnetization of nano-composites. The magnetization is due to induced circulatory currents on the periphery of these nano-composites in the superconducting state.



\end{document}